\documentclass[12pt]{iopart}

\usepackage{iopams}

\expandafter\let\csname equation*\endcsname\relax

\expandafter\let\csname endequation*\endcsname\relax

\usepackage{amsmath}
\usepackage{amsfonts}
\usepackage{amssymb}
\usepackage{hyperref}
\usepackage{graphicx}
\usepackage{color}
\usepackage{float}
\usepackage{wrapfig}
\usepackage{soul}
\usepackage{epsfig}
\usepackage{cite}
\newtheorem{thm}{Theorem}

\begin{document}

\title{Parameter and $q$ asymptotics of  $\mathfrak{L}_{q}$-norms of hypergeometric orthogonal polynomials}

\author{Jes\'us S. Dehesa}
\address{Instituto Carlos I de F\'{\i}sica Te\'orica y Computacional, Universidad de Granada, Granada 18071, Spain}
\address{Departamento de F\'{\i}sica At\'{o}mica, Molecular y Nuclear, Universidad de Granada, Granada 18071, Spain}
\vspace{10pt}
\ead{dehesa@ugr.es}
\author{Nahual Sobrino}
\address{Donostia International Physics Center, Paseo Manuel de Lardizabal 4, E-20018 San Sebasti\'an, Spain}
\address{Nano-Bio Spectroscopy Group and European Theoretical Spectroscopy Facility (ETSF), Departamento de Pol\'imeros y Materiales Avanzados: F\'isica, Qu\'imica y Tecnolog\'ia, Universidad del Pa\'is Vasco UPV/EHU, Avenida de Tolosa 72, E-20018 San Sebasti\'an, Spain}
\vspace{10pt}\ead{nahualcsc@dipc.org}

\begin{abstract}

 The three canonical families of the hypergeometric orthogonal polynomials (Hermite, Laguerre and Jacobi) control the physical wavefunctions of the bound stationary states of a great deal of quantum systems. The algebraic $\mathfrak{L}_{q}$-norms of these polynomials describe many physical, chemical and information-theoretical properties of these systems, such as e.g. the kinetic and Weizs\"acker energies, the position and momentum expectation values, the R\'enyi and Shannon entropies and the Cram\'er-Rao, the Fisher-Shannon and LMC measures of complexity. In this work we examine, partially review and solve the $q$-asymptotics and the parameter asymptotics (i.e., when the weight function's parameter tends towards infinity) of the unweighted and weighted $\mathfrak{L}_{q}$-norms for these  orthogonal polynomials. This study has been motivated by the application of these algebraic norms to the energetic, entropic and complexity-like properties of the highly-excited Rydberg and high-dimensional pseudo-classical states of harmonic (oscillator-like) and Coulomb (hydrogenic) systems, and other quantum systems subject to central potentials of anharmonic type.

\end{abstract}


\maketitle
\section{Introduction}

The hypergeometric orthogonal polynomials (HOPs) in one variable \cite{Szego1975,Chihara1978,Nikiforov1988,Ismail2005,Koekoek2010,Olver2010} have been used in numerous scientific areas ranging from applied mathematics, celestial mechanics and probability theory, to speech science, quantum mechanics and coding theory. This is basically because their mathematical structure has a rare combination of simplicity and usefulness.
In this paper we tackle, complement and partially review and solve the various asymptotics (degree, $q$ and weight-function parameter) of the integral functionals 
\begin{equation}
\label{eq:unweighted}
\mathcal{N}_{q}[p_{n}] := \int_{\Lambda}|p_{n}(x)|^{q}\,h(x)\,dx 
\end{equation}
and
 \begin{equation}\label{eq:weighted}
    W_q[p_n] := \int_\Lambda \left[p_n^2(x)\,h(x)\right]^q dx,
  \end{equation}
which are known (see e.g. \cite{Aptekarev1994,Aptekarev2010}) as the unweighted 
and weighted $\mathfrak{L}_{q}$-norms of the real hypergeometric polynomials $\{p_n(x)\}$, orthogonal with respect to the weight function $h(x)$ on the interval $\Lambda \subseteq{\mathbb{R}},$ respectively. They appear rather naturally in many branches of Mathematics and Physics. The three canonical families of HOPs are the Hermite polynomials $H_n(x)$, the Laguerre polynomials $L_n^{(\alpha)}(x),\, \alpha>-1,$ and the Jacobi polynomials $P_n^{(\alpha,\beta)}(x),\, \alpha,\beta>-1$. These norms are closely related to the entropy-like (Shannon \cite{Shannon1948}, R\'enyi \cite{Renyi1961,Renyi1970}) and complexity-like (Fisher-Shannon \cite{Romera2004,Angulo2008}, LMC-R\'enyi \cite{Pipek1997,Lopez2005,Lopezr2009,Sanchez2014}, Fisher-R\'enyi \cite{Romera2008,Antolin2009a,Romera2009,Antolin2009b,Toranzo2017rud,Puertas2017tor,Zozor2017}) measures of the Rakhmanov's density \cite{Rakhmanov1977} or probability density $\hat \rho_n(x)$ associated to the HOP $p_n(x)$, given by
\begin{equation}
\hat \rho_n(x) =  \hat{p}_n^2(x)\,h(x)= \frac{1}{\kappa_n} p_n^2(x)\,h(x), \label{eq:Rakhmanov}
\end{equation}
with the normalization constant $\kappa_n = \int_{\Lambda} |p_n(x)|^2 \, h(x) dx,$ and  where the symbol $\hat{p}_n(x) = p_n(x)/\kappa_n^\frac{1}{2}$ denotes the orthonormal polynomial. At times, the notation $\rho_{n}(x)=p_n^{2}(x)h(x)$ is also used.  \\

Mathematically, this density governs the asymptotics of the ratio of two polynomials with consecutive orders \cite{Rakhmanov1977} when the degree $n$ tends towards infinity. The algebraic norms (\ref{eq:unweighted}) and (\ref{eq:weighted}) quantify different configurational facets of the spread of the HOPs along the support interval $\Lambda$. They are, at times, much better probability estimators \cite{Sichel1947} than the ordinary moments $\nu_q[p_n] = \int_{\Lambda}x^n\,\rho_n(x)\,dx$; moreover, they are fairly efficient in the range where the ordinary moments are fairly inefficient \cite{Shenton1951,Romera2001,Romera2002}. Note that these algebraic norms are non-linear in probabilities and the feasible set of distributions which they define is non-convex \cite{Grendar2003}.  By increasing or decreasing its value, the $q$-parameter allows to enhance or diminish the contribution of the integrand over different regions to the whole integral. Higher values of $q$ make the function $[\rho_n(x)]^q$ to concentrate around the local maxima of the distribution, while the lower values have the effect of smoothing that function over its whole domain. It is in this sense that $q$ provides a powerful tool in order to get information on the structure of the probability density by means of the $\mathfrak{L}_{q}$-norms.\\
 
  Physically, the Rakhmanov's density describes the Born's probability density of the bound stationary states of numerous one and multidimensional quantum systems \cite{Nikiforov1988,Yanez1994,Dehesa2001,Dong2011,Brandon2013}. Then, the Rakhmanov's density may be often interpreted as the position and momentum density of single-particle quantum systems depending on the HOPs which control the system's wavefunctions in position and momentum states. So that the algebraic $\mathfrak{L}_{q}$-norms of the HOPs characterize different fundamental and/or experimentally measurable quantities of physical and chemical systems. In particular, these norms characterize the kinetic and Weizs\"acker energies \cite{Sears1980,Parr1989, Romera1994}, the position and momentum expectation values (see e.g. \cite{Assche2000}), the Heisenberg-like uncertainty relations \cite{Zozor2011} and numerous physical entropies and complexities of quantum systems with great scientific and technological interest \cite{Dehesa2011,Angulo2011}, such as e.g. the Shannon, R\'enyi and Tsallis entropies so that they are, in fact, the basic variables of the classical and quantum information theories \cite{Cover1991,Nielsen2000,Bruss2019}.\\

Up until now most analytical efforts on these algebraic norms have been addressed to bound them in many ways (see e.g. \cite{Borwein1995,Decarli2008,Guerrero2011,Grebenkov2013}), although some explicit expressions have been derived \cite{Dehesa2012,Sanchez2013,Toranzo2014, Puertas2018b,Dehesa2018}, and recently  reviewed \cite{Dehesa2021}, for the three canonical families of the real HOPs. However they are not easily  handy in the sense that, at times, they \textit{only} provide algorithmic expressions to compute them in a symbolic way because they require the evaluation of (a) Bessel polynomials of Combinatorics at the HOP expansion coefficients \cite{Comtet1974,Dehesa2012}, (b) some multivariate hypergeometric functions at unity (Jacobi case) or at $1/q$ (Hermite and Laguerre cases) \cite{Sanchez2013}, or (c) the logarithmic potential of the HOPs at the polynomial's zeros \cite{Dehesa2001,Aptekarev2010}. Numerically, the naive evaluation of the algebraic norms using quadratures is often not convenient due to the increasing number of integrable singularities when the polynomial degree $n$ is increasing, which spoils any attempt to achieve reasonable accuracy even for rather small $n$ (see e.g. \cite{Buyarov2004}). For the most complicated situations (i.e., when $n, q$ or the weight-function's parameter is very high) specific asymptotical approaches derived from approximation theory need to be developed \cite{Aptekarev1994,Aptekarev2010,Dehesa2021,Dehesa2014,Temme2017}. They are able to express the unweighted and weighted $\mathfrak{L}_{q}$-norms of the HOPs in a simple, transparent and compact form.\\

In this work we will update and, at times, solve the various asymptotics of the unweighted and weighted $\mathfrak{L}_{q}$-norms of the HOPs keeping in mind their close connection to the entropy and complexity-like quantities, and because of their relevance in the information theory of special functions and quantum systems and technologies \cite{Puertas2017b,Puertas2017,Dehesa2019,Dehesa2020}, as well as to facilitate their numerical and symbolic computation. The asymptotics of these algebraic norms for polynomials of degree $n$ ($n\rightarrow \infty$), weight-function's parameter  ($\alpha\rightarrow \infty$) and norm-parameter ($q\rightarrow \infty$) types have been previously considered and discussed in an incomplete form. The degree asympotics ($n\rightarrow \infty$) was initiated at the middle of the nineties in the seminal papers of Aptekarev et al \cite{Aptekarev1994,Aptekarev1994b,Aptekarev1996} and will not be considered here because it has been recently reviewed and discussed with some physical and mathematical applications in 2001 \cite{Dehesa2001} (see also \cite{Levin2003}), 2010 \cite{Aptekarev2010} and 2021 \cite{Dehesa2021}, respectively. The $q$-asymptotics ($q\rightarrow \infty$) for unweighted \cite{Toranzo2014} and weighted \cite{Dehesa2014} $\mathfrak{L}_{q}$-norms was tackled in 2014. The weight-function-parameter asymptotics ($\alpha\rightarrow \infty$) has been solved for the weighted norms of Laguerre and Gegenbauer polynomials  to a great extent by Temme et al \cite{Temme2017,Puertas2017} in 2017. The degree and the weight-function-parameter asymptotics have been recently used to evaluate the physical R\'enyi and Shannon entropies for the highly-excited (Rydberg) and high dimensional (pseudoclasical) states for quantum systems of harmonic (oscillator-like) \cite{Puertas2017b,Dehesa2017,Dehesa2019,Dehesa2020} and coulombian  (hydrogenic-like) types \cite{Dehesa2010,Toranzo2016,Toranzo2016b,Puertas2017,Dehesa2019}, as well as for some anharmonic potentials \cite{Dehesa2002a,Dehesa2002b}.  We do not consider here the norms of HOPs with varying weights (i.e., polynomials whose weight-function's parameter does depend on the polynomial degree), which are also of great mathematical and physical interest \cite{Buyarov1999,Levin2018,Aptekarev2021bis}.  \\

 This paper is structured as follows. We begin in Section \ref{relation} by briefly describing the relation of the algebraic norms (\ref{eq:unweighted}) and (\ref{eq:weighted}) to the entropic and complexity-like measures of the Rakhmanov's density of the HOPs.  In Section \ref{weightedLqasymptotics} we give the asymptotics ($q\rightarrow \infty$) for the weighted $\mathfrak{L}_{q}$-norms $W_q[p_n]$ of the HOPs.  In Section \ref{unweightedLq} we show the  asymptotics ($q\rightarrow \infty$) of the (unweighted) $\mathfrak{L}_{q}$-norms $\mathcal{N}_{q}(p_{n})$ of Jacobi polynomials $P_n^{(\alpha,\beta)}(x)$, and we point out that the corresponding norms for the Hermite $H_n(x)$ and Laguerre $L_n^{(\alpha)}(x)$ polynomials remain open.  In Section \ref{unweightedpar} we show the parameter asymptotics ($\alpha\rightarrow \infty$) of the (unweighted)  $\mathfrak{L}_{q}$-norms $\mathcal{N}_q[p_n]$ and the Shannon entropy $E[p_n]$ of the Laguerre, Jacobi  and Gegenbauer $C_n^{(\alpha)}(x)$ polynomials. In Section \ref{weightedpar} we find the parameter asymptotics ($\alpha\rightarrow \infty$) of the weighted  $\mathfrak{L}_{q}$-norms $W_q[p_n]$ of the Laguerre, Jacobi and Gegenbauer  polynomials, respectively. Finally, some concluding remarks are pointed out and a number of open related issues are identified in Section \ref{Conclud}.


\section{Relation to entropy and complexity-like measures of HOPs}
\label{relation}

In this section we briefly show the relationship of $\mathfrak{L}_{q}$-norms (\ref{eq:unweighted}) and (\ref{eq:weighted}) of the HOPs to the entropy-like measures (R\'enyi, Shannon) and complexity-like (LMC-R\'enyi, Fisher-R\'enyi, Fisher-Shannon) measures of their associated probability density or Rakhmanov density $\rho_n(x)$ given by Eq. (\ref{eq:Rakhmanov}). The R\'enyi \cite{Renyi1961,Renyi1970} and Shannon \cite{Shannon1948,Cover1991} entropies of the density $\rho_n(x)$ are defined by the expressions
   \begin{equation}
       R_q[\rho_n] =\frac{1}{1-q}\ln \int_{\Lambda} [\rho_n(x)]^q dx \equiv \frac{1}{1-q}\ln \mathcal{W}_q[\rho_n],\qquad q>0, \quad q\neq 1,
   \end{equation}
and
  \begin{equation}\label{ShaRen}
     S[\rho_n] = \lim_{q\rightarrow 1} R_q[\rho_n]=-\int_{\Lambda} \rho_n(x) \ln \rho_n(x) dx,
  \end{equation}
respectively. Now, by keeping in mind (\ref{eq:weighted}), one has that the weighted norms of the HOPs  are $W_q[p_n] = \mathcal{W}_q[\rho_n]$. Then, the R\'enyi entropies \cite{Renyi1970} of the HOP $p_n(x)$  are related to the weighted $\mathfrak{L}_{q}$-norms as 
\begin{equation}\label{renyinorm}
       R_q[p_n] = \frac{1}{1-q} \ln W_q[p_n],
   \end{equation}
  with $q>0$ and $q\neq1$. They quantify numerous $q$-dependent configurational aspects of the spreading of the density $\rho_n(x)$ over the support  $\Lambda$. When $q \to  1$ the R\'enyi entropies tend towards the Shannon-like integral functional $S[p_n]$, which measures the total spreading of $\rho_n(x)$. So, this functional is the limiting case 
\begin{equation}\label{shannonfun}
     S[p_n] = \lim_{q\rightarrow 1} R_q[p_n]=-\int_{\Lambda} \rho_n(x) \ln \rho_n(x) dx := S[\rho_n] =E[p_n]+ I[p_n],
  \end{equation}
 and correspondingly
\begin{equation}
	S[\hat p_n] = -\int_{\Lambda} \frac{1}{\kappa_n}p_n^2(x)\,h(x)\,\ln \left[\frac{1}{\kappa_n}p_n^2(x)\,h(x)\right]  dx 
	= \ln \kappa_n + \frac{1}{\kappa_n}\left(E[p_n] + I[p_n]\right)
\end{equation}
with the polynomial functionals
\begin{equation}\label{IntegralS}
I[p_n]:= -\int_{\Lambda} p_n^2(x)h(x)\,\ln h(x)  dx
\end{equation}
and
\begin{equation}\label{eq:integralE}
E[p_n]:= -\int_{\Lambda} p_n^2(x)h(x)\,\ln p_n^2(x)  dx.
\end{equation}
The functional $I[p_n]$ have been explicitly determined \cite{Sanchez2000} by means of the coefficients of the second-order differential equation of the HOPs. However, the explicit determination of the functional $E[p_n]$ in terms of the degree and the parameters of the weight function $h(x)$ is a formidable task, not yet solved for the HOPs  except (a) for the Chebyshev polynomials of the first and second type and for some Gegenbauer polynomials \cite{Dehesa1994,Dehesa2001}, (b) in some extreme cases: when  ($n\rightarrow \infty$) and when the parameters of the weight $h(x)$ go towards $\infty$, and (c) for Laguerre polynomials by means of a somewhat highbrow expression which involve the evaluation of a bivariate Appell function of second kind $F_{A}^{(2)}(x_{1},,x_{2})$ and a multivariate Lauricella function $F_{A}^{(r)}(x_{1},\ldots,x_{r})$ evaluated at unity and its qth-derivative (see \cite[Eq. (23)]{Toranzo2020}). This functional $E[p_n]$, usually called Shannon entropy of the HOPs $p_n(x)$, can be expressed in terms of the unweighted $\mathfrak{L}_{q}$-norms as
\begin{equation}\label{EPN2}
	E\left[p_n\right] = \left.2 \frac{d \mathcal{N}_{q}\left[p_n\right]}{d q} \right|_{q=2},
\end{equation}
for orthogonal polynomials, and
\begin{align}\label{ShaDer}
E[\hat{p}_n]&= - \lim_{q\rightarrow 1} \frac{1}{q-1} \ln  \int |\hat{p}_n(x)|^{2q} h(x) dx 
 = - \lim_{q\rightarrow 1} \frac{\partial}{\partial q} \mathcal{N}_{2q}[\hat{p}_n],
\end{align}
for orthonormal polynomials.
In addition, it is also fulfilled \cite{Angulo2000} that 
\begin{equation}\label{eq:SdW}
S[\rho_n]=-\left. \frac{d \mathcal{W}_q[\rho_n]}{d q} \right|_{q=1}.
\end{equation}

The (biparametric) LMC-R\'enyi complexity measure \cite{Pipek1997,Lopez2005,Lopezr2009,Sanchez2014} of the Rakhmanov's density $\rho_n(x)$ defined as
\begin{equation}
 \label{eq:6new}
 C_{\alpha,\beta}[\rho_n]: = e^{R_{\alpha}[\rho_n]-R_{\beta}[\rho_n]}, \quad 0<\alpha <\beta<\infty, \quad \alpha,\beta \neq 1,
 \end{equation}
 can be expressed in terms of the weighted norms as
 \begin{equation}
 	C_{\alpha,\beta}[\rho_n] = \left(\mathcal{W}_{\alpha}[\rho_n]\right)^{\frac{1}{1-\alpha}} \times \left(\mathcal{W}_{\beta}[\rho_n]\right)^{\frac{-1}{1-\beta}}.
 \end{equation}
 This quantity extends a number of other measures such as the shape-R\'enyi complexity \cite{Antolin2009b} given by $C_{\alpha,2}[\rho_n] =  e^{R_{\alpha}[\rho_n]} \times  \mathcal{W}_2[\rho_n]$, and the plain LMC (Lopez Ruiz-Mancini-Calbet) complexity \cite{Lopez1995,Catalan2002} given by $C_{1,2}[\rho_n] =  e^{S[\rho_n]} \times  \mathcal{W}_2[\rho_n]$ which measures the combined balance of the  deviation of $\rho_n$  from the equilibrium or disequilibrium (as given by $\mathcal{W}_2[\rho_n] = e^{-R_2[\rho_n]}$) and its total extent over the density support (as given by the Shannon entropy power or Shannon length $\mathcal{L}_1^S[p_n] = e^{S[p_n]}$ \cite{Hall1999}.\\

The Fisher-Shannon complexity of the polynomial $p_n(x)$ is given \cite{Angulo2008,Romera2004} by
\begin{equation}
     \label{fishershannon}
      \mathcal{C}_{FS}[p_n]=F[p_n] \times \frac{1}{2 \pi e} e^{2 S[p_n]}=\frac{1}{2 \pi e} F[p_n] \times \left(\mathcal{L}_{1}^S[p_n]\right)^2,
  \end{equation}
where the symbols $S[p_n]$ and $F[p_n]$ denote the Shannon-like entropic functional of the polynomial $p_n(x)$ given by (\ref{shannonfun}) and the Fisher information \cite{Fisher1925,Frieden2004} of the Rakhmanov density $\rho_n(x)$ associated to $p_n(x)$ defined as
 \begin{equation*}
     F\left[\rho_n\right]=\int_{\Lambda} \frac{[\rho'_n(x)]^2}{\rho_n(x)}dx,
  \end{equation*}
  respectively. Opposite to the R\'enyi and Shannon entropies, the Fisher information has a local character because it is a functional of the derivative of $\rho_n(x)$, what allows it to be explicitly determined for all the HOPs in terms of  the degree and the weight-function's parameters. This has been done for the first time from the second-order differential equation of HOPS \cite{Sanchez2005} (see also \cite{Yanez2008,Dehesa2012}). \\
  
 The natural generalization of the Fisher-Shannon measure is the Fisher-R\'enyi complexity \cite{Romera2008,Antolin2009a,Romera2009,Antolin2009b,Toranzo2017rud,Puertas2017tor,Zozor2017}, which is defined by 
 \begin{equation}
     \label{fisherrenyi}
      \mathcal{C}_{FR}[p_n]=F[p_n] \times \frac{1}{2 \pi e} e^{2 R_q[p_n]}=\frac{1}{2 \pi e} F[p_n] \times \left(\mathcal{L}_{q}^R[p_n]\right)^2,
  \end{equation}
 where the symbol $\mathcal{L}_{q}^R[p_n]$ denotes the R\'enyi entropy power or R\'enyi length \cite{Hall1999} of the HOP $p_n(x)$ given by  
  \begin{equation}\label{eq:renyi_length}
    \mathcal{L}_q^R[\rho_n] = e^{R_q[\rho_n]}=\left(\mathcal{W}_q[\rho_n]\right)^{\frac{1}{1-q}}=\left\{\int_\Lambda \left[\rho_n(x)\right]^q dx\right\}^{\frac{1}{1-q}}.
   \end{equation}
   Note that the Shannon length is the limiting case of the R\'enyi length since 
  \begin{equation}\label{eq:shannon_length}
      \mathcal{L}_1^S[\rho_n] = \lim_{q\rightarrow 1}  \mathcal{L}_q^R[\rho_n]=e^{S[\rho_n]}=e^{-\int_\Lambda \rho_n(x) \ln\rho_n(x) dx}, 
  \end{equation}
The entropy-like quantities ($R_q[\rho_n]$, $S[\rho_n]$, $F[\rho_n]$) are complementary because they grasp  different single spreading facets of the probability density $\rho(x)$. The R\'enyi and
Shannon entropies are measures of the various aspects of the extent to which the density is in fact concentrated, and the Fisher information is a quantitative estimation of the 
oscillatory character of the density since it estimates the pointwise concentration of the probability over its support interval $\Lambda$. The three complexity measures $(C_{FS}[\rho_n], C_{FR}[\rho_n], C_{\alpha,\beta}[\rho_n])$, which are dimensionless, quantify different two-fold configurational facets of the spread of the HOPs along the support interval. They are known to be invariant under translation and scaling transformation \cite{Yamano2004a,Yamano2004b}, universally bounded from below by unity \cite{Dembo1991,Guerrero2011,Angulo2012,Lopez2009a}, and monotonic \cite{Rudnicki2016}.
\\

In the next sections we will determine the previously defined weighted and unweighted norms $(W_q[p_n], \mathcal{N}_q[p_n])$ of the HOPs $\{p_n(x), \mathrm{deg}\, p_n=n\}$, which control the entropy- and complexity-like properties of such polynomials over the orthogonality support interval $\Lambda$. These polynomials are orthogonal with respect to the weight function $h(x)$ on the interval $\Lambda \in (a,b) \subseteq{\mathbb{R}},$ so that \cite{Nikiforov1988,Olver2010}
\begin{equation}
\int_{\Lambda} p_n(x) p_m(x)\, h(x) dx = \kappa_n\, \delta_{n,m},\label{eq:orthogonality_relation}
\end{equation}
where the weight function $h(x)$ has the expressions
\begin{align}
h^{H}(x)&=e^{-x^2}; \quad 
h^{L}_{\alpha}(x)=x^\alpha e^{-x}; \quad  
h^{J}_{\alpha,\beta}(x)=(1-x)^\alpha (1+x)^\beta,
\end{align}
for the three canonical HOPs families of Hermite $H_n(x), x\in(-\infty,+\infty)$, Laguerre $L_n^{(\alpha)}(x),\alpha>-1, x\in[0,+\infty)$, and Jacobi $P_n^{(\alpha,\beta)}(x), (\alpha,\beta>-1), x\in [-1,+1]$ types, respectively. The corresponding normalization constants are
\begin{gather}
	\kappa^{H}_n = \sqrt{\pi}\, n!\,\, 2^n; \quad
	\kappa^{L}_{n,\alpha} = \Gamma(n+\alpha+1)/n!; \quad  \text{and} \nonumber\\
	\quad \kappa^{J}_{n,\alpha,\beta} =\frac{2^{\alpha+\beta+1}\Gamma(\alpha+n+1)
\Gamma(\beta+n+1)}{n!(\alpha+\beta+2n+1)\Gamma(\alpha+\beta+n+1)}, 
\label{eq:ctes}
\end{gather}
respectively. The special Jacobi case $\alpha=\beta=\lambda-\frac{1}{2}$ corresponds to the ultraspherical or Gegenbauer polynomials $C_n^{(\lambda)}(x),\lambda> -\frac{1}{2},\,\lambda\ne 0$ with slightly different normalization (see e.g. \cite{Olver2010}); so that its weight function $h^{G}_{\lambda}(x)=(1-x^2)^{\lambda-\frac{1}{2}}$ and the corresponding normalization constant is $\kappa^{G}_{n,\lambda} = \frac{2^{1-2\lambda} \pi \Gamma(n+2\lambda)}
{\left[\Gamma(\lambda) \right]^2(n+\lambda)n!}$. Note that $\kappa_n=1$ for the orthonormal polynomials $\hat{p}_n(x)$ of Hermite $\hat{H}_n(x)$, Laguerre $\hat{L}_n^{(\alpha)}(x)$ and Jacobi $\hat{P}_n^{(\alpha,\beta)}(x)$ types.

\section{Weighted $\mathfrak{L}_{q}$-norms $W_q[p_n]$  of HOPs. Asymptotics ($q\rightarrow \infty$)}
\label{weightedLqasymptotics}

The weighted norms $W_q[p_n]$ of the three canonical HOPs families (Hermite, Laguerre, Jacobi) can be evaluated for all $n$ by the two following analytical/algorithmic approaches: using the multivariate Bell polynomials of Combinatorics \cite{Comtet1974,Dehesa2012,Toranzo2014} when $q\in\mathbb{N}$, and by means of some multivariate hypergeometric functions evaluated \cite{Sanchez2013} (see also a recent review in section 5 of \cite{Dehesa2021}) at unity and at $1/q$, or by determining the logarithmic potential of these polynomials evaluated at their zeros \cite{Dehesa2001,Aptekarev2010}. However these approaches are, at times, very computationally demanding, especially for high values of the degree $n$, the norm-parameter $q$ and the weight-function parameter(s). Then, it is almost mandatory to tackle both asymptotics ($n\rightarrow \infty$) and ($q\rightarrow \infty$), and the asymptotics associated to the the weight-function parameter(s). The degree asymptotics of HOPs has been solved and recently reviewed \cite{Aptekarev1994,Aptekarev2010,Dehesa2021} as already said. The weight-function-parameter asymptotics will be analyzed later on.\\

The purpose of this section is to show and discuss the asymptotics $(q \rightarrow \infty)$ for the weighted norms $W_q[p_n]$ of the three canonical families of the real HOPs $\{p_n(x)\}$, which are defined by (\ref{eq:weighted}). To do it we use the Laplace's method, obtaining \cite{Wong1989,Dehesa2014} that
 \begin{eqnarray}\label{eq:weighted2}
    W_q[p_n] :&= \int_\Lambda \left[p_n^2(x)\,h(x)\right]^q dx = \int_\Lambda e^{q f(x)}\,dx\nonumber\\
    \label{eq:weighted3}
     &= e^{q f(x_0)} \left[ \sqrt{\frac{2 \pi}{-q\, f''(x_0)}} +\mathcal{O}(q^{-1})\right], \quad q \to \infty,
  \end{eqnarray}
  where $x_0=x_0(n)$, which denotes the value of the abcissa at which the absolute maximum of the function $f(x)= \ln h(x)+\ln p_n^2(x)$ is achieved, is given by the solution of the equation
  \begin{equation}
     \label{ratio}
     \frac{p'_n(x_0)}{p_n(x_0)}=-\frac{1}{2} \frac{h'(x_0)}{h(x_0)}.
  \end{equation}
  So, this asymptotics is basically controlled by the extremum $x_0$.

\subsection{Hermite polynomials}


In this case the absolute maximum $x_0$ is given by the equation 
\[
     x_0H_n(x_0)=2nH_{n-1}(x_0).
\]
and the second derivative $f''_H(x_0)$ has the value
\[
     f''_H(x_0)=2x_0^2-4n-2.
\]
Then, according to Eq. (\ref{eq:weighted2}) we obtain that the weighted norms of the Hermite polynomials fulfill the asymptotics
 \begin{align}
     \label{asymptotic_hermite}
      W_q[H_n] &=  \int_{-\infty}^{+\infty} \left[h^H(x) H_n^2(x)\right]^q \,dx\nonumber\\
      &= 2 \left[h^H(x_0) H_n^2(x_0)\right]^q
     \left[
     \sqrt{\frac{2\pi}{q(4n-2x_0^2+2)}}+\mathcal{O}(q^{-1})
     \right],\,  q \to \infty.
  \end{align}

  For $n=0$, one has $H_0(x) = 1$ and $x_0 =0$ so that this asymptotical formula gives the exact value $\sqrt{\frac{\pi}{q}}$. For $n=1$ one has $H_1(x) = 2x$ and $x_0=1$, so that the asymptotical value of the corresponding weighted norm is $2^{2q+1} e^{-q} 
     \left[
     \sqrt{\frac{\pi}{2q}}
     +\mathcal{O}(q^{-1})
     \right]$.
 Moreover, for $n=2$, we have that $H_2(x) = 4x^2-2$ and $x_0 = \sqrt{\frac{5}{2}}$, so that the weighted norm of the corresponding polynomial has the asymptotical value
  \begin{equation}
     \label{asymptoticvalues_hermite}  
     \quad 2^{6q+1} e^{-\frac{5}{2}q}
     \left[
     \sqrt{\frac{2 \pi}{5q}}
     +\mathcal{O}(q^{-1})
     \right].
  \end{equation}

  \subsection{Laguerre polynomials}


In this case, according to Eq. (\ref{ratio}), the absolute maximum $x_0=x_0(n)$ is  given by  \begin{equation}
     \label{equation_laguerre}
     \left(\frac{\alpha}{x_0}-1\right) L_n^{(\alpha)}(x_0)=2 L_{n-1}^{(\alpha+1)}(x_0),
  \end{equation}
 and the second derivative $f''_L(x_0)$ has the value
\[
     f''_L(x_0)=\frac{\alpha^2}{2 x_0^2}-\frac{2n+\alpha+1}{x_0}+\frac{1}{2}.
\]
Then, according to (\ref{eq:weighted2}) we obtain the following asymptotics for the weighted norm of Laguerre polynomials $L_n^{(\alpha)}(x)$ 
   \begin{align}
     \label{asymptotic_laguerre}
    W_q[L_n^{(\alpha)}] &=\int_{0}^{+\infty} \left[h^{L}_{\alpha}(x) \left[L_n^{(\alpha)}(x)\right]^2\right]^q \,dx \nonumber \\
     &= \left[h^{L}_{\alpha}(x_0) \left[L_n^{(\alpha)}(x_0)\right]^2\right]^q \times \left[
     \sqrt{\frac{2\pi}{q\left(-\frac{\alpha^2}{2x_0^2}+\frac{2n+\alpha+1}{x_0}-\frac{1}{2}\right)}}
     +\mathcal{O}(q^{-1})
     \right],
  \end{align}
for $q \to +\infty$ and $\alpha>0$. For the particular cases $n=0; 1$, one has $L_0^{(\alpha)}(x)=1; L_1^{(\alpha)}(x)=\alpha+1-x$ and the absolute maximum values $x_0(0)=\alpha; x_0(1)= \frac{1}{2}\left(2\alpha+3-\sqrt{8\alpha+9}\right)$, respectively. Then, the weighted norms of the corresponding Laguerre polynomials have the asymptotical values
 \[
     \alpha^{q  \alpha} e^{-q  \alpha} 
     \left[
     \sqrt{\frac{2 \pi \alpha}{q}}
     +\mathcal{O}(q^{-1})
     \right] \quad {\rm and}
\quad \left[x_0^{\alpha} e^{-x_0} (1+\alpha-x_0)^2\right]^q
     \left[
     \sqrt{\frac{2 \pi}{-q f''_L(x_0)}}
     +\mathcal{O}(q^{-1})
     \right]\\
   \]
respectively, with
    \[
      f''_L(x_0)=
      \frac{3\sqrt{8\alpha+9}-8\alpha-9}{\left(\sqrt{8\alpha+9}-2\alpha-3\right)^2}.
	\]

\subsection{Jacobi polynomials}


In this case, according to Eq. (\ref{ratio}), the absolute maximum $x_0=x_0(n)$ is  given by 
 \begin{equation}
      \label{ratio_jacobi}
      \frac{P_{n-1}^{(\alpha+1,\beta+1)}(x_0)}{P_{n}^{(\alpha,\beta)}(x_0)}=-\frac{1}{\alpha+\beta+n+1}
      \left(\frac{-\alpha}{1-x_0}+\frac{\beta}{1+x_0}\right),
   \end{equation}
 and the second derivative $f''_J(x_0)$ has the value    
 \begin{align}
       f''_J(x_0)&=
-\left(\alpha+\frac{\alpha^2}{2}\right) \frac{1}{(1-x_0)^2}-\left(\beta+\frac{\beta^2}{2}\right)\frac{1}{(1+x_0)^2} - \frac{\alpha\beta}{1-x_0^2}
       \nonumber\\&-\frac{2n(n+\alpha+\beta+1)}{1-x_0^2}
       +\frac{\beta-\alpha-(\alpha+\beta+2)x_0}{1-x_0^2}\left[\frac{\beta}{1+x_0}-\frac{\alpha}{1-x_0}\right].
              \label{second_derivative_jacobi}
    \end{align}
    Then, according to (\ref{eq:weighted2}) we obtain the following asymptotics for the weighted norm of Jacobi polynomials $P_n^{(\alpha,\beta)}(x), (\alpha,\beta>-1), x\in [-1,+1]$
    \begin{align}
     \label{asymptotic_jacobi}
        W_q[P_n^{(\alpha,\beta)}] &= \int_{-1}^{+1} \left[h^{J}_{\alpha,\beta}(x) \left[P_n^{(\alpha,\beta)}(x)\right]^2\right]^q \,dx
     \nonumber\\
     &= \left[h^{J}_{\alpha,\beta}(x_0) \left[P_n^{(\alpha,\beta)}(x_0)\right]^2\right]^q
     \left[
     \sqrt{\frac{2\pi}{-q f''_J(x_0)}}
     +\mathcal{O}(q^{-1})
     \right]
  \end{align}
for $ q\to \infty$ and $\alpha,\beta>0$. Finally, in the particular case where $n=0$, $\alpha>0$ and $\beta>0$ we can find from Eq. (\ref{ratio_jacobi}) and 
(\ref{second_derivative_jacobi}) that 
   \[
      x_0=\frac{\beta-\alpha}{\alpha+\beta} \quad {\rm and} \quad f''_J(x_0)=-\frac{(\alpha+\beta)^3}{4 \alpha \beta},
   \]
respectively. Then, from Eq. (\ref{asymptotic_jacobi}) with these values of $x_0$ and $f''_J(x_0)$ we obtain the following value
\begin{align}
     W_q[P_0^{(\alpha,\beta)}] &=\int_{-1}^{+1} \left[h^{J}_{\alpha,\beta}(x) \left(P_0^{(\alpha,\beta)}(x)\right)^2\right]^q dx \nonumber\\
   &=
     2^{q(\alpha+\beta)} \left(\frac{\alpha}{\alpha+\beta}\right)^{\alpha q} \left(\frac{\beta}{\alpha+\beta}\right)^{\beta q}
     \left[
     \sqrt{\frac{8 \pi \alpha \beta}{q (\alpha+\beta)^3}}
     +\mathcal{O}(q^{-1})
     \right]
\end{align}
for the leading term of the asymptotics ($q \to \infty$) of $P_0^{(\alpha,\beta)}(x)=1$.\\

 \section{$\mathfrak{L}_{q}$-norms $\mathcal{N}_{q}[p_{n}]$ of HOPs. Asymptotics ($q\rightarrow \infty$). Open problems}
\label{unweightedLq}

The unweighted norms (\ref{eq:unweighted})  of the three canonical HOPs families (Hermite, Laguerre, Jacobi)  can be evaluated for all $n$ by using the combinatorial Bell polynomials \cite{Comtet1974,Toranzo2014} when $q=2k$ and $k\in\mathbb{N}$. Indeed, they can be expressed as
\begin{equation}
\label{eq:11}
\mathcal{N}_{q}[p_{n}]= \int_{\Lambda}|p_{n}(x)|^{q}\,h(x)\,dx =\sum_{t=0}^{nq} \frac{q!}{(t+q)!}B_{t+q,q}(c_{0}, 2! c_{1}, \ldots, (t+1)! c_{t})\, \mu_{t}
\end{equation}
where $c_j$ denotes the coefficients of the power expansion $p_{n}(x)=\sum_{k=0}^{n} c_{k}x^{k}$, and the $B$-symbol denotes the multivariate Bell polynomials given by 
\begin{equation}
B_{m,l}(c_{1},c_{2},\ldots, c_{m-l+1})
=\sum_{\pi(m,l)}\frac{m!}{j_{1}!j_{2}!\ldots j_{m-l+1}!}\left(\frac{c_{1}}{1!}\right)^{j_{1}}\left(\frac{c_{2}}{2!}\right)^{j_{2}}\ldots \left(\frac{c_{m-l+1}}{(m-l+1)!}\right)^{j_{m-l+1}},\nonumber\\
\label{eq:8}
\end{equation}
where the sum runs over all partitions $\pi(m,l)$ such that $j_{1}+j_{2}+\ldots+j_{m-l+1}=l$ and $j_{1}+2j_{2}+\ldots+(m-l+1)j_{m-l+1}=m$. Moreover, $\mu_{t}$ denotes the moment of order $t$ of the weight function $h(x)$; i.e.
\begin{equation}
\label{eq:12}
\mu_{t}=\int_{\Lambda} x^{t}\, h(x)\, dx, \hspace{0.5cm} t=0,1,\ldots
\end{equation}
whose values are known to be 
\begin{align}
\label{eq:17} \mu_{2t+1}[H]&=0 ,\,\, \mu_{2t}[H]=\Gamma\left(t+\frac{1}{2}\right); \quad
 \mu_{t}[L]=\Gamma(1+\alpha+t) \\
\mu_{t}[J]&=\Gamma(1+t)\Big[(-1)^{t}\frac{\Gamma(1+\beta)}{\Gamma(2+t+\beta)}{}_2 F_1(-\alpha,t+1;2+t+\beta;-1) \nonumber \\
&    +\frac{\Gamma(1+\alpha)}{\Gamma(2+t+\alpha)}{}_2 F_1(-\beta,t+1;2+t+\alpha;-1)\Big]\label{eq:19}
\end{align}
for Hermite, Laguerre and Jacobi polynomials, respectively. Then, the expressions (\ref{eq:11})-(\ref{eq:19}), together with the expansion coefficients $c_j$  (see e.g. \cite{Olver2010}), provide an algorithmic procedure to determine the unweighted $\mathcal{N}_{q}$ norms (\ref{eq:unweighted}) of the Hermite, Laguerre and Jacobi polynomials in terms of $q, n$ and the parameter of the corresponding weight function (see section $2$ of \cite{Toranzo2014} for further details). Alternatively, the unweighted quantities $\mathcal{N}_{q}[p_{n}]$  can be also obtained by using the  Srivastava-Niukkanen linearizing formulas \cite{Sanchez2013,Dehesa2018} of powers of Laguerre and Jacobi polynomials, already employed for the calculation of the weighted norms. The corresponding results, however, require the evaluation at unity of some multivariate hypergeometric functions Lauricella type. These two approaches to find both symbolically and numerically the unweighted norms $N_{q}[p_n]$ of the HOPs are computationally demanding, especially in the (qualitatively different) extremal cases: $q\rightarrow \infty$, $n\rightarrow \infty$ and when the parameters of the weight function become very large. In such cases it is more convenient to use specific asymptotical approaches derived from approximation theory \cite{Lubinsky1988,Wong1989,Temme2015,Aptekarev2012}.\\ 

In this section we tackle and discuss the asymptotics $q\rightarrow \infty$ for the (unweighted) $\mathfrak{L}_{q}$-norms $\mathcal{N}_{q}[P_n^{(\alpha,\beta}]$ of Jacobi polynomials by means of the Laplace method \cite{Wong1989}. Unfortunately, this method is not applicable to Hermite and Laguerre polynomials, as it is explained later on. Therefore, the asymptotics $q\rightarrow \infty$ of the following (unweighted) $\mathfrak{L}_{q}$-norms   
\begin{equation}
\label{eq:3bis}
\mathcal{N}_{q}[H_{n}] = \int_{\Lambda}|H_{n}(x)|^{q}\,h^{H}(x)\,dx  = \int_{-\infty}^{+\infty}e^{-x^2}\, |H_n(x)|^{q} dx
\end{equation}
and
\begin{equation}
\label{eq:3}
\mathcal{N}_{q}[L_{n}^{(\alpha)}] = \int_{\Lambda}|L_{n}^{(\alpha)}(x)|^{q}\,h^{L}_{\alpha}(x)\,dx =  \int_{0}^{+\infty} x^\alpha e^{-x}\,|L_{n}^{(\alpha)}(x)|^{q}\,dx
\end{equation}
remains open for the future. \\

The evaluation of the unweighted norms of HOPs in the other two extremal situations, i.e., when $n\rightarrow \infty$ and when $\alpha\rightarrow \infty$, are also relevant problems not yet solved. This problem appear, however, in numerous chemical and physical problems related to the highly-excited or Rydberg (i.e., when $n\rightarrow \infty$) and  the high dimensional or quasi-classical (i.e. when $\alpha\rightarrow \infty$)  quantum states of harmonic and coulombian systems; indeed, their wavefunctions are controlled by Hermite and Laguerre polynomials for one and multidimensional cases and in both position and momentum spaces, respectively.

Let us now show the evaluation of the unweighted norms of the Jacobi polynomials for the extremal case $q\rightarrow \infty$.

\subsection{Asymptotics ($q\rightarrow \infty$) for the $\mathfrak{L}_{q}$-norms  of Jacobi polynomials}

In this section we determine the asymptotics ($q\rightarrow \infty$) for the unweighted $\mathfrak{L}_{q}$-norms $\mathcal{N}_{q}[P_n^{(\alpha,\beta)}]$ of the Jacobi polynomials, defined by
\begin{equation} \label{eq:unwJacobi}
\mathcal{N}_{q}[P_n^{(\alpha,\beta)}]= \int_{\Lambda}|P_n^{(\alpha,\beta)}(x)|^{q}\,h^{J}_{\alpha,\beta}(x)\,dx = \int_{-1}^{1} (1-x)^\alpha (1+x)^\beta \left|P^{(\alpha,\beta)}_{n}(x)\right|^{q} dx,
\end{equation}
The asymptotic behaviour ($q\rightarrow \infty$) of the unweighted $L_{q}$-norms $\mathcal{N}_{q}[p_n]$ of the polynomials $p_n(x)$ given by Eq. (\ref{eq:unweighted}), can be evaluated by the  extended Laplace method (see Theorem 1 of \cite[Chapter 2]{Wong1989}, and \cite[section 4]{Toranzo2014}). However, this method demands the existence of a global maximum of the function $|p_{n}(x)|$. Then, it is not applicable to Hermite and Laguerre polynomials because the functions $|H_n(x)|$ and $|L_n^{(\alpha)}(x)|$ do not have such maximum in the intervals of orthogonality $(-\infty,+\infty)$ and $(0,+\infty)$, respectively. Now, for the Jacobi polynomials $P_n^{(\alpha,\beta)}(x)$ the maximum is achieved at $x=-1$ if $\beta\ge\alpha>-1$, $\beta\ge-\frac12$ \cite[Eq. 18.14.2]{Olver2010}, and at $x=1$ if $\alpha\ge\beta>-1$, $\alpha\ge-\frac12$  \cite[Eq. 18.14.1]{Olver2010}, with the values
\begin{equation}
|P_n^{(\alpha,\beta)}(-1)|=\frac{(\beta+1)_n}{n!}; \qquad |P_n^{(\alpha,\beta)}(1)|=\frac{(\alpha+1)_n}{n!}
\label{eq:maximum_m1}
\end{equation}
Now, to obtain the unweighted norms (\ref{eq:unwJacobi}) we use the first order asymptotics ($q\rightarrow \infty$) 
\begin{equation}
\int_a^b \phi(x)e^{-q t(x)}dx=e^{-q t(a)} \left(
\Gamma\left(\frac{\gamma}{\mu}\right)\frac{b_0}{\mu a_0^{\gamma/\mu}} q^{-\frac{\gamma}{\mu}} 
+ O\left(q^{-\frac{1+\gamma}{\mu}}\right)
\right).
\label{eq:iq_wong}
\end{equation}
where the functions $t(x)>t(a), \forall x\in(a,b),$ and $\phi(x)$  have the expansions
\[
t(x)= t(a)+a_0(x-a)^\mu +\cdots,\qquad \phi(x)= b_0(x-a)^{\gamma-1} +\cdots.
\]
Then, for Jacobi polynomials we have that $\phi(x)=(1-x)^\alpha(1+x)^\beta$ and $t(x)=-\ln\left|P_n^{(\alpha,\beta)}(x)\right|$. Now, let us consider first the case when $\beta\ge\alpha>-1$, $\beta\ge-\frac12$; so, according to Eq. (\ref{eq:maximum_m1}), the maximum occurs at $x=a=-1$, fulfilling the requirement of the Laplace method. Thus, we obtain the expansions
\[
\phi(x)=2^\alpha (x+1)^{\beta}+\cdots
\]
so that $b_0=2^\alpha$, $\gamma=\beta+1$, and
\[
t(x)=-\ln\left|P_n^{(\alpha,\beta)}(-1)\right|
-\frac12(n+\alpha+\beta+1)\frac{P_{n-1}^{(\alpha+1,\beta+1)}(-1)}{P_n^{(\alpha,\beta)}(-1)}(x+1)+\cdots,
\]
so that $\mu=1$, and
\[
a_0=-\frac12(n+\alpha+\beta+1)\frac{P_{n-1}^{(\alpha+1,\beta+1)}(-1)}{P_n^{(\alpha,\beta)}(-1)}
=
\frac12 (n+\alpha+\beta+1)\frac{n}{\beta+1}.
\]
The substitution of these values of $a_{0}$, $b_{0}$, $\gamma$ and $\mu$ in Eq. (\ref{eq:iq_wong}) gives rise to the following values \cite{Toranzo2014} for the unweighted norms of Jacobi polynomials
\begin{equation}
 \mathcal{N}_{q}[P_n^{(\alpha,\beta)}]=\left(
\frac{(\beta+1)_n}{n!}
\right)^q \left(2^\alpha \Gamma(\beta+1) \left(
\frac{2(\beta+1)}{(n+\alpha+\beta+1)n}
\right)^{\beta+1}q^{-\beta-1}
+O\left(q^{-\beta-2}\right)
\right),
\label{eq:asymp_m1}
\end{equation}
if $\beta\ge\alpha>-1$, $\beta\ge-\frac12$. Similarly, with the change of variable $x\to -x$, the unweighted norms of Jacobi polynomials have the values
\begin{equation}
\mathcal{N}_{q}[P_n^{(\alpha,\beta)}]=\left(
\frac{(\alpha+1)_n}{n!}
\right)^q \left(2^\beta \Gamma(\alpha+1) \left(
\frac{2(\alpha+1)}{(n+\alpha+\beta+1)n}
\right)^{\alpha+1}q^{-\alpha-1}
+O\left(q^{-\alpha-2}\right)
\right),
\label{eq:asymp_1}
\end{equation}
if $\alpha\ge\beta>-1$, $\alpha\ge-\frac12$. Note the simplicity and transparency of expressions (\ref{eq:asymp_m1}) and (\ref{eq:asymp_1}), valid for large $q$, with respect to the general expressions (\ref{eq:11})-(\ref{eq:19}) which, although valid for all $q$, are somewhat highbrow, not analytically handy.\\

\section{$\mathfrak{L}_{q}$-norms $\mathcal{N}_{q}[p_{n}]$ and Shannon entropy $E\left[p_n\right]$  of HOPs. Parameter asymptotics ($\alpha\rightarrow \infty$).}
\label{unweightedpar}

The unweighted $\mathfrak{L}_{q}$-norms (\ref{eq:unweighted}) of the three parameter-dependent HOPs families (Laguerre, Jacobi, Gegenbauer) can be explicitly evaluated, as mentioned above, although in a not so handy way because their analytical expressions require the evaluation of some multivariate hypergeometric functions in an algorithmic form. The latter is specially true when the parameter(s) of their weight function has large values. Rarely, they can be determined recursively such as for the Gegenbauer polynomials \cite{Assche2000}. Then, it is mandatory to develop some asymptotical approaches derived from approximation theory to determine these algebraic norms in a simple and transparent way. The asymptotics ($n\rightarrow \infty$) of the algebraic norms was already solved in the seminal work of Aptekarev et al \cite{Aptekarev1994} (see also the review \cite{Dehesa2021}). \\

The \textit{leitmotiv} of this section is the asymptotics ($\alpha\rightarrow \infty$) of the $\mathfrak{L}_{q}$-norms $\mathcal{N}_{q}[p_{n}]$ and the Shannon entropy $E\left[p_n\right]$ of Laguerre, Jacobi and Gegenbauer polynomials. We first update the existing approaches for the asymptotics ($\alpha\rightarrow \infty$) of the algebraic norms $\mathcal{N}_q(L_n^{(\alpha)})$ and $\mathcal{N}_q(P_n^{(\alpha,\beta)})$ of Laguerre and Jacobi polynomials, given by Eqs. (\ref{eq:3}) and (\ref{eq:unwJacobi}), respectively. Then, according to Eq. (\ref{EPN2}), we calculate from these quantities the Shannon entropies (\ref{eq:integralE}) given by the  expressions
\begin{equation}\label{eq:EPN2Lag}
	E\left[L_n^{(\alpha)}\right] :=-\int_{0}^{\infty}
\left[L_n^{(\alpha)}(x)\right]^2 h^{L}_{\alpha}(x) \ln
\left[L_n^{(\alpha)}(x)\right]^2dx = \left.2 \frac{d \mathcal{N}_{q}\left[L_n^{(\alpha)}\right]}{d q} \right|_{q=2},
\end{equation}
\begin{equation}\label{eq:EPN2Jac}
	E\left[P_n^{(\alpha,\beta)}\right] :=-\int_{-1}^{+1}
\left[P_n^{(\alpha,\beta)}(x)\right]^2 h^J_{\alpha,\beta}(x) \ln
\left[P_n^{(\alpha,\beta)}(x)\right]^2dx = \left.2 \frac{d \mathcal{N}_{q}\left[P_n^{(\alpha,\beta)}\right]}{d q} \right|_{q=2},
\end{equation}
and
\begin{align}\label{eq:shaGegen}
E\left[C_n^{(\lambda)}\right]&=-\int_{-1}^{+1}
\left[C_n^{(\lambda)}(x)\right]^2 h^{G}_{\lambda}(x) \ln
\left[C_n^{(\lambda)}(x)\right]^2dx =2 \frac{d}{dq}\left[\mathcal{N}_{q}\left[C_n^{(\lambda)}\right]\right]_{q=2}
\end{align}
for Laguerre, Jacobi and Gegenbauer polynomials, respectively. Physically, these entropic quantities describe the Shannon entropies of the high-dimensional quantum states of numerous quantum systems, such as e.g. the $D$-dimensional oscillator-like and hydrogenic systems (see e.g. \cite{Dehesa2019,Toranzo2020}). Basically, this is because the wavefunctions of these systems are controlled by the Laguerre and Gegenbauer polynomials, $L_n^{(\alpha)}(x)$ and $C_n^{(\lambda)}(x)$, where the parameters $\alpha$ and $\lambda$ are linear functions of the space dimensionality $D$ of the system (see e.g. \cite{Toranzo2020}). \\

\subsection{$\mathfrak{L}_{q}$-norms $\mathcal{N}_q(L_n^{(\alpha)})$ and Shannon entropy $E\left[L_n^{(\alpha)}\right]$ of Laguerre polynomials. Parameter asymptotics.}
\label{parasymunweiLaguerre} 


 To obtain the asymptotics of the unweighted $\mathcal{N}_q(L_n^{(\alpha})$ norm and the Shannon entropy $E[L_n^{(\alpha}]$ of the Laguerre polynomials $L_n^{(\alpha}(x)$, given by Eqs. (\ref{eq:3}) and (\ref{eq:EPN2Lag}) respectively, we use the following theorem of Temme et al \cite{Temme2017} and its extension (see \cite[section 5]{Temme2017}). This recent result allows one to evaluate the general entropy-like functionals of Laguerre polynomials $I_{1} (m,\alpha)$ and $I_{2} (m,\alpha)$ given below, which include the wanted functionals $\mathcal{N}_q(L_n^{\alpha})$ and $E[L_n^{(\alpha}]$ as  particular cases.\\

\begin{thm}\cite{Temme2017}\label{prop:P1}
Let $\alpha, \lambda, q$, and $\mu$ be positive real numbers, and $m$ a positive natural number. Then, the unweighted functional of Laguerre polynomials
\begin{equation}
\label{eq:Lag19}
I_{1} (m,\alpha)= \int\limits_{0}^{\infty}x^{\mu-1}e^{-\lambda x} \left|\mathcal{L}_{m}^{(\alpha)}(x)\right|^{q}\,dx, 
\end{equation}
fulfills the asymptotic expansion 
\begin{equation}
\label{eq:Lag20}
 I_{1}(m,\alpha)\sim   \frac{\alpha^{q m}\Gamma(\mu)}{\lambda^\mu (m!)^q} \sum_{k=0}^{\infty}\frac{D_{k} }{\alpha^k},\quad \alpha \to\infty, \text{and rest of parameters fixed.}
 \end{equation}
The first coefficients are
\begin{equation}
\label{eq:Lag21}
D_{0}=1,\quad D_1=\frac{q m (-2\mu+m\lambda+\lambda)}{2\lambda}, 
\end{equation}
and
\begin{equation}
\label{eq:Lag22}
\begin{array}{@{}r@{\;}c@{\;}l@{}}
D_2&=&
q m\bigl(-12\mu \lambda q m^2+24\mu \lambda -12\mu \lambda q m-4m^2\lambda ^2-6m\lambda ^2+
3m^3\lambda ^2q \\ [8pt]
&&-12\mu ^2+12\mu ^2q m-12\mu +12\mu q m+6\lambda ^2q m^2-2\lambda ^2+3\lambda ^2q m\bigr)/(24\lambda ^2).
\end{array}
\end{equation}
\end{thm}
From these expressions we obtain
\begin{equation}
\label{eq:Lag19}
\int_{0}^{\infty}x^{\mu-1}e^{-\lambda x} \left|L_{m}^{(\alpha)}(x)\right|^{\kappa}\,dx \sim   \frac{\alpha^{\kappa m}\Gamma(\mu)}{\lambda^\mu (m!)^\kappa}, \qquad \alpha \to\infty.\,\text{and rest of parameters fixed.} 
\end{equation}
Moreover, by differentiating the expansion \eqref{eq:Lag20} with respect to $q$ and taking $q=2$ afterwards, we find that the generalized Shannon-like integrals $I_2(m,\alpha)$ defined by
\begin{equation}
\label{eq:intro02}
I_{2}(m,\alpha)= \int_{0}^{\infty} x^{\mu-1}e^{-\lambda x}\bigl(\mathcal{L}^{(\alpha)}_{m}(x)\bigr)^{2}\ln\bigl(\mathcal{L}^{(\alpha)}_{m}(x)\bigr)^{2}\, dx,  
\end{equation}
have the following values
 \begin{equation}
\label{eq:Lag23}
I_2(m,\alpha)=2\left.\frac{\partial}{\partial q}I_1(m,\alpha)\right\vert_{q=2}\sim \frac{\alpha^{2 m}\Gamma(\mu)}{\lambda^\mu (m!)^2} 
\bigl(\ln\frac{\alpha^{2m}}{(m!)^2}\sum_{k=0}^{\infty}\frac{D_{k} }{\alpha^k}+2\sum_{k=0}^{\infty}\frac{D_{k}^\prime }{\alpha^k}\bigr),
\end{equation}
for $\alpha \to\infty$ and the rest of parameters are fixed. The derivatives $D'_k$ are with respect to $q$.\\

Furthermore, let us now consider the extension (see \cite[section 5]{Temme2017}) of the previous theorem for the case $\mu= O(\alpha)$ in the special form $\mu=\sigma+\alpha$, $\lambda =1$ and with $\sigma$ a fixed real number. Then, we can use the limit (see \cite[Eqn.~18.7.26]{Olver2010})
\begin{equation}
\label{eq:more20}
\lim_{\alpha\to\infty}\left(\frac{2}{\alpha}\right)^{\frac12m}\mathcal{L}_{m}^{(\alpha)}\left(\sqrt{2\alpha}\,x+\alpha\right)=\frac{(-1)^m}{m!}H_m(x),
\end{equation}
so that we have the asymptotic relation
\begin{equation}
\label{eq:more21}
\mathcal{L}_{m}^{(\alpha)}(\alpha x)\sim \left(\frac{\alpha}{2}\right)^{\frac12m}\frac{(-1)^m}{m!}H_m\left(\sqrt{\frac{\alpha}{2}}(x-1)\right).
\end{equation}
Then, we obtain in the first approximation that
\begin{equation}
\label{eq:more22}
\int\limits_{0}^{\infty}x^{\alpha+\sigma-1}e^{- x} \left|\mathcal{L}_{m}^{(\alpha)}(x)\right|^q\,dx\sim\alpha^{\alpha+\sigma} e^{-\alpha}\frac{1}{(m!)^q}\left(\frac{\alpha}{2}\right)^{\frac12 q m}
\int\limits_{-\infty}^{\infty}e^{-\frac12\alpha y^2} \left|H_m\left(\sqrt{\frac{\alpha}{2}}\,y\right)\right|^{q}\,dy,
\end{equation}
when $\alpha \to\infty$ and the rest of parameters ($\sigma, \lambda = 1, q, m$) are fixed. This expression can be alternatively found and rewritten \cite{Belega2017} as 
\begin{equation}
\label{eq:more23}
\int\limits_{0}^{\infty}x^{\alpha+\delta}e^{- x} \left|\mathcal{L}_{m}^{(\alpha)}(x)\right|^q\,dx \sim c_{m,q} \left(\frac{\alpha}{e}\right)^\alpha \, \alpha^{\delta+(mq+1)/2},\qquad \alpha \to\infty	
\end{equation}
  with 
\begin{equation}
	c_{m,q} = \frac{\mathcal{N}_{q}[H_{m}]}{(m!)^q\, 2^{mq-1/2}},
\end{equation}
being $m$ a positive integer number, $\delta$ a real number and q a positive real number, and $\mathcal{N}_{q}[H_{m}]$ the unweighted $\mathfrak{L}_{q}$-norm of Hermite polynomials defined by Eq. (\ref{eq:3bis}). The constant $c_{m,q}$, which does not depend on $\alpha$, is controlled by the unweighted norm of the Hermite polynomials which can be explicitly found for all $m$ (see e.g. \cite{Dehesa2018}) and in the limit $m \to \infty$ (see \cite{Aptekarev2012}).  From this asymptotical expression and an identity similar to (\ref{eq:EPN2Lag}), we obtain the following parameter asymptotics for the extended Shannon entropic functional
\begin{equation}
	\int\limits_{0}^{\infty}x^{\alpha+\delta}e^{- x} \left|\mathcal{L}_{m}^{(\alpha)}(x)\right|^2\,\ln \left|\mathcal{L}_{m}^{(\alpha)}(x)\right|^2\, dx \sim \frac{\sqrt{2\,\pi}}{(m-1)!}\left(\frac{\alpha}{e}\right)^\alpha \, \alpha^{\delta+m+1/2}\ln\alpha,\quad \alpha \to\infty	.
\end{equation}
Finally, putting $\delta=1$ we have from the last two asymptotical expressions the parameter asymptotics 
\begin{equation}
	\mathcal{N}_{q}[L_{n}^{(\alpha)}] :=  \int_{0}^{+\infty} x^\alpha e^{-x}\,|L_{n}^{(\alpha)}(x)|^{q}\,dx \sim c_{m,q} \left(\frac{\alpha}{e}\right)^\alpha \, \alpha^{(mq+1)/2},\qquad \alpha \to\infty
\end{equation}
for the unweighted norms of Laguerre polynomials, and
\begin{equation}
	E[L_{n}^{(\alpha)}] := \int\limits_{0}^{\infty}x^{\alpha}e^{- x} \left|\mathcal{L}_{m}^{(\alpha)}(x)\right|^2\,\ln \left|\mathcal{L}_{m}^{(\alpha)}(x)\right|^2\, dx \sim  \frac{\sqrt{2\,\pi}}{(m-1)!}\left(\frac{\alpha}{e}\right)^\alpha \, \alpha^{m+3/2}\ln\alpha,\quad \alpha \to\infty	 
\end{equation}
for the Shannon entropy of Laguerre polynomials\cite{Belega2017,Dehesa2021b}.

\subsection{$\mathfrak{L}_{q}$-norms and Shannon entropy of Jacobi and Gegenbauer polynomials. Parameter asymptotics.}
\label{parasymunweiJacobi}

To obtain the parameter asymptotics ($\alpha\rightarrow \infty, \beta\, \text{fixed}$) of the unweighted norm $\mathcal{N}_q(P_n^{(\alpha,\beta)})$  and the Shannon entropy $E\left[\hat{P}_n^{(\alpha,\beta)}\right]$ of the Jacobi polynomials, given by Eqs. (\ref{eq:unwJacobi}) and (\ref{eq:EPN2Jac}) respectively, we follow the lines of Sobrino et al \cite[subsect. 3.2]{Sobrino2021}. First, from Eq. (\ref{eq:unwJacobi}) and the limiting relation
\begin{equation}\label{eq:limitJac}
	\lim_{\alpha \rightarrow \infty} \frac{P_n^{(\alpha,\beta)}(x)}{P_n^{(\alpha,\beta)}(1)} = \left(\frac{1+x}{2}\right)^{n}, \qquad \text{with}\qquad P_n^{(\alpha,\beta)}(1) = \frac{\Gamma(\alpha+n+1)}{n!\,\Gamma(\alpha+1)},
\end{equation}
we find the asymptotics
\begin{align}\label{eq:Np_norm_2_s}
\mathcal{N}_{p}\left[P_n^{(\alpha,\beta)}\right]&\sim   \frac{\Gamma(\alpha+n+1)}{n!}
\frac{\Gamma(1+np+\beta)}{\Gamma(2+\alpha+np+\beta)}2^{1+\alpha+\beta}; \quad \alpha\rightarrow \infty, \beta \, \text{fixed}
\end{align}
Thus, according to Eqs. (\ref{EPN2}) and (\ref{eq:Np_norm_2_s}), one has that the asymptotics of the Shannon entropy  of the orthogonal Jacobi polynomials $P_n^{(\alpha,\beta)}(x)$ is given as
\begin{align}
E\left[P_n^{(\alpha,\beta)}\right] &:= -\int_{-1}^{+1} (1-x)^\alpha (1+x)^\beta
\left[P_n^{(\alpha,\beta)}(x)\right]^2  \ln
\left[P_n^{(\alpha,\beta)}(x)\right]^2dx\nonumber\\
	&  \sim 2^{2+\alpha+\beta}\alpha^{-n-\beta-1}\left(\frac{\Gamma(1+2n+\beta)}{\Gamma(n)}(\psi(1+2n+\beta)-\ln(\alpha))+\mathcal{O}(\alpha^{-2})\right) \nonumber\\
		\label{eq:shaJac}
\end{align}
when $\alpha \rightarrow \infty, \beta\, \text{fixed}$ and being $\psi(x)= \frac{\Gamma^{'}(x)}{\Gamma(x)}$ the digamma function .

A similar result follows for $\beta\to\infty$ by exchanging $\alpha\leftrightarrow\beta$. The explicit expression of these entropies is not yet known \cite{Dehesa2021}, although their asymptotical behavior when $n \rightarrow \infty$ is controlled \cite{Aptekarev1994b,Aptekarev1996,Aptekarev2010}.\\

From the last two asymptotical expressions (\ref{eq:Np_norm_2_s}) and (\ref{eq:shaJac}) with $\alpha=\beta = \lambda-1/2$ and taking into account the following relation
\begin{equation}
  	C_n^{(\lambda)}(x) =c_{n,{\lambda}}P_{n}^{(\lambda-\frac{1}{2},\lambda-\frac{1}{2})}(x)\equiv \frac{\Gamma (\lambda+\frac{1}{2})}{\Gamma(2\lambda)}\,\frac{\Gamma(n+2\lambda)}{\Gamma(n+\lambda+\frac{1}{2})}\,P_{n}^{(\lambda-\frac{1}{2},\lambda-\frac{1}{2})}(x),
  \end{equation}
one can obtain the asymptotics $(\lambda \rightarrow \infty)$  of the $\mathfrak{L}_{q}$-norms $\mathcal{N}_{q}[C_{n}^{(\lambda)}]$ and the Shannon entropy $E\left[C_n^{(\lambda)}\right]$ of Gegenbauer polynomials, respectively. These entropies have not yet been explicitly evaluated for all $(n,\lambda)$ except for integer $\lambda$, but their asymptotical behavior when $n\rightarrow \infty$ has been determined \cite{Buyarov2000,Dehesa2001,Vicente2007}.

\subsubsection{Parameter asymptotics for $\mathfrak{L}_{q}$-norms $\mathcal{N}_q(C_n^{(\lambda)})$ and Shannon entropy $E\left[C_n^{(\lambda)}\right]$ of Gegenbauer polynomials}.\\


The interest on the asymptotics ($\lambda\rightarrow \infty$) of the Gegenbauer polynomial themselves and their algebraic norms has been a long standing problem \cite{Elbert1992,Dehesa2001,Decarli2005,Buyarov2000,Vicente2007,Temme2017,Puertas2017,Toranzo2020,Dehesa2021b} because of fundamental and quantum applications; this is basically because the Gegenbauer polynomials control the angular part of the quantum wavefunctions of central potentials in position space and the momentum wavefunctions of Coulomb systems (see e.g. the reviews \cite{Dehesa2019,Dehesa2021c,Dehesa2021d}.\\

So, let us center around the asymptotics ($\lambda\rightarrow \infty$) of the unweighted $\mathfrak{L}_{q}$-norms of orthogonal Gegenbauer polynomials given by
\begin{align}\label{Np_norm}
\mathcal{N}_{q}\left[C_n^{(\lambda)}\right] :=\int_{-1}^{1}h^{G}_{\lambda}(x)\left|C_n^{(\lambda)}\right|^{q}dx,
\end{align}
and the Shannon entropy (\ref{eq:shaGegen}), where $h^{G}_{\lambda}(x) = (1-x^{2})^{\lambda-\frac{1}{2}}$. Then, we take into account the limiting relation
\begin{equation}\label{eq:limitGeg}
	\lim_{\lambda \rightarrow \infty} \frac{C_n^{(\lambda)}(x)}{C_n^{(\lambda)}(1)} = x^n,\qquad \text{with}\qquad C_n^{(\lambda)}(1) = \frac{(n+2\lambda-1)!}{n! \,(2\lambda-1)!},
\end{equation}
to obtain \cite{Dehesa2021b}
\begin{align}\label{Np_norm_2}
\mathcal{N}_{q}\left[C_n^{(\lambda)}\right]\sim \left[C_n^{(\lambda)}(1)\right]^{q}\frac{\Gamma(\frac{1}{2}(1+nq))\Gamma(\frac{1}{2}+ n)}{\Gamma(1+\lambda+\frac{nq}{2})} \sim \frac{\Gamma(\frac{1}{2}(1+nq))}{n!^q}, \qquad \lambda\rightarrow \infty.
\end{align}
And for the orthonormal Gegenbauer polynomials $\hat{C}_n^{(\lambda)}(x)= C_n^{(\lambda)}(x)\,(\kappa_{n,\lambda}^{G})^{-\frac{1}{2}}$, we have the following asymptotics
\begin{align}\label{Np_norm_2}
\mathcal{N}_{q}\left[\hat{C}_n^{(\lambda)}\right]& = \frac{1}{(\kappa_{n,\lambda}^{G})^{q/2}}\mathcal{N}_{q}\left[C_n^{(\lambda)}\right]  \sim \left[\frac{C_n^{(\lambda)}(1)}{(\kappa_{n,\lambda}^{G})^{1/2}}\right]^{q}\frac{\Gamma(\frac{1}{2}(1+nq))\Gamma(\frac{1}{2}+{\color{magenta}n})}{\Gamma(1+\lambda+\frac{nq}{2})}\nonumber\\
&	\sim \frac{\Gamma(\frac{1}{2}(1+nq))}{n!^q}\left(\frac{n!^{\frac{q}{2}}\lambda^{q}}{\pi^{\frac{q}{4}}}+\mathcal{O}(\lambda^{-\frac{q}{4}})\right), \qquad \lambda\rightarrow \infty,
\end{align}
of the corresponding unweighted norms.

Finally, according to (\ref{eq:shaGegen}) and (\ref{Np_norm_2}), one has that the Shannon entropy  of the orthogonal Gegenbauer polynomials fulfills the asymptotics
\begin{align}
	E\left[C_n^{(\lambda)}\right] &:=\int_{-1}^{+1}
\left[C_n^{(\lambda)}(x)\right]^2 h^{G}_{\lambda}(x) \ln
\left[C_n^{(\lambda)}(x)\right]^2dx \nonumber\\
	&\sim 2\,\kappa_{n,\lambda}^{G} \left(\ln \left[\frac{(n+2\lambda-1)!}{n! \,(2\lambda-1)!}\right] + \frac{n}{2}\, \psi(\frac{2n+1}{2})-\frac{n}{2}\,\psi(n+2\lambda+1)\right),\nonumber \\
\end{align}
with the normalization constant $\kappa^{G}_n = \frac{2^{1-2\lambda} \pi \Gamma(n+2\lambda)}
{\left[\Gamma(\lambda) \right]^2(n+\lambda)n!}$. And for the orthonormal polynomials $\hat{C}_n^{(\lambda)}(x)$, we have the parameter asymptotics
\begin{gather}
	E\left[\hat{C}_n^{(\lambda)}\right] \sim 2\left(\ln \left[\frac{(n+2\lambda-1)!}{n! \,(2\lambda-1)!}\right] + \frac{n}{2}\,\psi(\frac{2n+1}{2})-\frac{n}{2}\,\psi(n+2\lambda+1)\right)
\sim 2\ln\left(\frac{\lambda^{n}2^{n}}{n!}\right)
	\label{E_lambda_infinity}
\end{gather}
in a simple and elegant form.\\

\section{Weighted $\mathfrak{L}_{q}$-norms $W_q[p_n]$  of HOPs. Parameter asymptotics}
\label{weightedpar}

This section is devoted to the parameter asymptotics ($\alpha\rightarrow \infty$) for the weighted $\mathfrak{L}_{q}$-norms of the three parameter-dependent HOPs families of Laguerre, Jacobi and Gegenbauer types defined by Eq. (\ref{eq:weighted}) and denoted by $W_q[L_{n}^{(\alpha)}]$, $W_q[P_n^{(\alpha,\beta)}]$ and $W_q[C_n^{(\lambda)}]$, respectively. These integral functionals have been of great mathematical interest in the theory of trigonometric series and extremal polynomials since Bernstein's times \cite{Bernstein1954,Suetin1979,Lubinsky1988,Zygmund2002}. More recently, they have been shown to be explicitly evaluated, as mentioned above, although in a highbrow, not so handy way because the associated analytical expressions require the evaluation of either the multivariate Bell polynomials so useful in combinatorics or some multivariate hypergeometric functions of Lauricella or Srivastava-Daoust types in an algorithmic form \cite{Srivastava1988,Srivastava2003,Sanchez2013,Comtet1974}. This is specially so when the parameter(s) of their weight function has large values. Then, it is mandatory to develop some asymptotical approaches derived from approximation theory to determine these algebraic norms.\\

 Physically, the asymptotical values of the weighted $\mathfrak{L}_{q}$-norms for the Laguerre, Jacobi and Gegenbauer polynomials provide various energy-dependent quantites and the R\'enyi, Shannon and Tsallis entropies of the high-dimensional pseudo-classical states of a great deal of quantum systems of harmonic and Coulomb types (e.g. the dimensional oscillator- and hydrogenic-like systems) in a simple and transparent way. The latter is basically because the corresponding wavefunctions of these systems are controlled by the mentioned HOPs where the parameter of their weight functions is directly dependent on the space dimensionality. \\

\subsection{Weighted $\mathfrak{L}_{q}$-norms $W_q(L_n^{(\alpha)})$ of Laguerre polynomials. Parameter asymptotics.}
\label{parasymweiLaguerre}

The parameter asymptotics ($\alpha\rightarrow \infty$) for the weighted $\mathfrak{L}_{q}$-norms $W_q(L_n^{(\alpha)})$ of (orthogonal) Laguerre polynomials defined by
\begin{equation}
	W_q[L_n^{(\alpha)}] = \int_{0}^{\infty}
\left(\left[L_n^{(\alpha)}(x)\right]^2 h^{L}_{\alpha}(x)\right)^q\,dx = \int_{0}^{\infty} \,x^{q\alpha}\,e^{-q\alpha}\,\left[L_n^{(\alpha)}(x)\right]^{2q}\,dx,
\end{equation}
can be determined by \eqref{eq:Lag20} and \eqref{eq:Lag19} derived from Theorem 1 of Temme et al \cite{Temme2017}. Then, with the values $\mu= q\alpha+1,\, \lambda=q$ and $\kappa = 2q$, this general asymptotical formula provides the required asymptotics for $W_q[L_n^{(\alpha)}]$:
\begin{equation}\label{omega2L}
	W_q[L_n^{(\alpha)}] \sim \frac{\alpha^{2qn}\,\Gamma(q\alpha+1)}{q^{q\alpha+1}\,(n!)^{2q}}, \qquad \alpha \to\infty.
\end{equation}
Moreover, the weighted $\mathfrak{L}_{q}$-norms $W_q(\hat{L}_n^{(\alpha)})$ of orthonormal Laguerre polynomials fulfill the parameter asymptotics
\begin{equation}\label{omega2Lbis}
	W_q[\hat{L}_n^{(\alpha)}] = \frac{1}{(\kappa_{n,\alpha}^{L})^q}\, W_q[L_n^{(\alpha)}] \sim \frac{1}{(\kappa_{n,\infty}^{L})^{q}} \frac{\alpha^{2qn}\,\Gamma(q\alpha+1)}{q^{q\alpha+1}\,(n!)^{2q}}, \qquad \alpha \to\infty,
\end{equation}
with
\begin{equation}
	\kappa^{L}_{n,\infty} = \lim_{\alpha \rightarrow \infty} \kappa^{L}_{n,\alpha} \sim \frac{\sqrt{2\pi}}{n!} \left(\frac{\alpha}{e}\right)^{\alpha}\,\alpha^{n+1/2}, \qquad \alpha\rightarrow \infty,
	\label{kappa_constant_L}
\end{equation}
where we have taken into account that the normalization constant $\kappa^{L}_{n,\alpha}$ is given by Eq. \eqref{eq:ctes}, and keeping in mind that $\Gamma(z) \sim e^{-z}\,z^z\,\left(\frac{2\pi}{z} \right)^{1/2}$ (see Eq. 5.11.3 of \cite{Olver2010}), one has that 
\begin{equation}\label{omega2Ltris}
	W_q[\hat{L}_n^{(\alpha)}] \sim \frac{\alpha^{q(n-\frac{1}{2})+\frac{1}{2}}}{\sqrt{q}(n!)^{q}(2\pi)^{\frac{1}{2}(q-1)}}, \qquad \alpha \to\infty,
\end{equation}
which extends to all $q$ the following asymptotics
\begin{equation}\label{omega2L}
	W_2[\hat{L}_n^{(\alpha)}] =\alpha^{2n}\left(\frac{1}{2\,(n!)^{2}\sqrt{\pi \alpha}}+\mathcal{O}(\alpha^{-3/2})\right), \qquad \alpha \to\infty
\end{equation}
recently found (see Eq. 32 of \cite{Dehesa2021b}) for the second order norm $W_2[\hat{L}_n^{(\alpha)}]$, which is a fundamental ingredient of the LMC complexity of the orthonormal Laguerre polynomials.

\subsection{Weighted $\mathfrak{L}_{q}$-norms of Jacobi and Gegenbauer polynomials. Parameter asymptotics.}
\label{parasymweiJacobi} 

In this section we show the parameter asymptotics ($\alpha\rightarrow \infty, \beta\, \text{fixed}$) for the weighted $\mathfrak{L}_{q}$-norms 
\begin{equation}\label{eq:omega2Jac}
	W_q[P_n^{(\alpha,\beta)}] = \int_{-1}^{+1}
\left(\left|P_n^{(\alpha,\beta)}(x)\right|^2 h_{\alpha,\beta}(x)\right)^q\,dx = \int_{-1}^{+1} \,(1-x)^{q\alpha} (1+x)^{q\beta}\,\left|P_n^{(\alpha,\beta)}(x)\right|^{2q}\,dx,
\end{equation}
of (orthogonal) Jacobi polynomials $P_n^{(\alpha,\beta)}(x)$, and the parameter asymptotics ($\lambda\rightarrow \infty$) for the corresponding norms 
  \begin{equation}\label{eq:omega2Geg}
	W_q[C_n^{(\lambda)}] = \int_{-1}^{+1}
\left(\left|C_n^{(\lambda)}(x)\right|^2 h^{G}_{\lambda}(x)\right)^q\,dx = \int_{-1}^{+1} \,(1-x^2)^{q\lambda-q/2}\,\left|C_n^{(\lambda)}(x)\right|^{2q}\,dx,
\end{equation}
of (orthogonal) Gegenbauer polynomials $C_n^{(\lambda)}(x)$.\\

To obtain the parameter asymptotics ($\alpha\rightarrow \infty, \beta\, \text{fixed}$) of the weighted norm $W_q(P_n^{(\alpha,\beta)})$ of the Jacobi polynomials $P_n^{(\alpha,\beta)}(x)$, we use the limiting relation (\ref{eq:limitJac}) in Eq. (\ref{eq:omega2Jac}), obtaining the asymptotics 
\begin{align}\label{eq:Np_norm_2_simplified}
W_{q}\left[P_n^{(\alpha,\beta)}\right]\sim&\left[ P_n^{(\alpha,\beta)}(1)\right]^{2q}4^{-nq}\left(\frac{1}{1+2nq+q\beta}{}_{2}F_{1}(1,-q\alpha,2+2nq+q\beta,-1)\right.\nonumber\\
&\left.+\frac{1}{1+q\alpha}{}_{2}F_{1}(1,-q(2n+\beta),2+q\alpha,-1)\right), 
\quad \alpha\rightarrow \infty, \beta \, \text{fixed}\nonumber\\
& \sim\left[P_n^{(\alpha,\beta)}(1)\right]^{2q}  \frac{2^{1+q(\alpha+\beta)}\Gamma(1+q\alpha)\Gamma(1+2nq+q\beta)}{\Gamma(2+q(\alpha+\beta+2n))},
\quad \alpha\rightarrow \infty, \beta \, \text{fixed}\nonumber\\
\end{align}
which generalizes to all $q$ the asymptotics given by (Eq. 35 of \cite{Sobrino2021}) for the second-order norm $W_{q}\left[P_n^{(\alpha,\beta)}\right]$ of the orthogonal Jacobi polynomials. Moreover, the weighted $\mathfrak{L}_{q}$-norms $W_q(\hat{P}_n^{(\alpha,\beta)})$ of orthonormal Jacobi polynomials fulfill the parameter asymptotics 
\begin{equation}\label{omega2Lbis}
	W_q[\hat{P}_n^{(\alpha,\beta)}] = \frac{1}{(\kappa_{n,\alpha,\beta}^{J})^q}\, W_q[P_n^{(\alpha,\beta)}]  \sim  \frac{2^{1-q}}{(n!)^{q}q^{1+q(\beta+2n)}}\frac{\Gamma(1+2nq+n\beta)}{\Gamma(\beta+n+1)}\alpha^{q-1}, \quad \alpha\rightarrow \infty, \beta \, \text{fixed}
\end{equation}
which extends to all $q$ the asymptotics
\begin{equation}\label{W2oG}
	W_2[\hat{P}_n^{(\alpha,\beta)}] = \frac{1}{(\kappa_{n,\alpha,\beta}^{J})^2}\,W_2[P_n^{(\alpha,\beta)}] \sim \frac{\Gamma(1+4n+2\beta)}{2^{2(1+2n+\beta)}\,(n!)^{2}\,\Gamma(1+n+\beta)}\alpha ,\qquad \alpha\rightarrow \infty, \beta \, \text{fixed}
\end{equation} 
recently found (see Eq. 36 of \cite{Sobrino2021}) for the second-order norm $W_2[\hat{P}_n^{(\alpha,\beta)}]$, which is a fundamental ingredient for the measure of complexity of the orthonormal Jacobi polynomials.

Finally, to obtain the parameter asymptotics ($\lambda\rightarrow \infty$) of the weighted norm $W_q(C_n^{(\lambda)})$ of the Gegenbauer polynomials $C_n^{(\lambda)}(x)$, we follow a similar procedure. We use the limiting relation (\ref{eq:limitGeg}) in Eq. (\ref{eq:omega2Geg}), obtaining the asymptotics 
\begin{align}\label{eq:Np_norm_2_simplified}
W_{q}\left[C_n^{(\lambda)}\right]&\sim    \left[C_n^{(\lambda)}(1)\right]^{2q}\left[\frac{(1+(-1)^{2nq})\Gamma(\frac{1}{2}+nq)\Gamma(1+q(\lambda-\frac{1}{2}))}{2\Gamma(\frac{3}{2}+q(n+\lambda-\frac{1}{2}))}\right]   \nonumber\\
& \sim (1+(-1)^{2nq})\Gamma(\frac{1}{2}+nq)\frac{2^{2nq}}{q^{\frac{1}{2}+nq}(n!)^{2q}}   \lambda^{nq-\frac{1}{2}}, \quad \lambda\rightarrow \infty.
\end{align}

Moreover, the weighted $\mathfrak{L}_{q}$-norms $W_q(\hat{C}_n^{(\lambda)})$ of orthonormal Laguerre polynomials fulfill the parameter asymptotics 
\begin{equation}\label{eq:omega2Lbis}
	W_q[\hat{C}_n^{(\lambda)}] = \frac{1}{(\kappa_{n,\lambda}^{G})^q}\, W_q[C_n^{(\lambda)}] \sim    (1+(-1)^{2nq})\frac{2^{nq-1}\Gamma(\frac{1}{2}+nq)}{q^{\frac{1}{2}+nq}\pi^{\frac{q}{2}}(n!)^{q}}\lambda^{\frac{1}{2}(q-1)}, \qquad \lambda \to\infty,
\end{equation}
where we have also taken into account that $\kappa_{n,\lambda}^{G}\sim \lambda^{n-1/2}2^{n}\sqrt{\pi}/n!$ when $\lambda\rightarrow \infty$; and for $q=2$, this result simplifies as
\begin{equation}\label{W2oG}
	W_2[\hat{C}_n^{(\lambda)}] = \frac{1}{(\kappa_{n,\lambda}^{G})^2}\,W_2[C_n^{(\lambda)}] \sim \frac{\Gamma(\frac{1}{2}+2n)}{\sqrt{2}\pi (n!)^{2}}\lambda^{\frac{1}{2}}, \qquad \lambda \rightarrow \infty.
\end{equation}
Remark that  the last two expressions (\ref{eq:Np_norm_2_simplified}) and (\ref{eq:omega2Lbis}) extend to all $q$ the corresponding algebraic norms for the orthogonal and orthonormal Gegenbauer polynomials obtained by Eq. (65) and (66) of \cite{Dehesa2021b}, respectively. \\


\section{Conclusions} \label{Conclud}

In this work, the present knowledge of the spreading of the hypergeometric orthogonal polynomials (HOPs) is examined and updated by means of the unweighted and weighted $\mathfrak{L}_{q}$-norms, given by Eqs. (\ref{eq:unweighted}) and (\ref{eq:weighted}) respectively. Emphasis is placed on the three possible asymptotics of these algebraic norms: the degree asymptotics, the $q$ asymptotics and the weight-function parameter asymptotics. The latter two asymptotics are partially reviewed and solved. This study has been physically motivated by the applications of these norms to the energetic, entropic and complexity-like properties of the highly-excited Rydberg and high-dimensional pseudo-classical states of harmonic (oscillator-like) and Coulomb (hydrogenic) systems.\\

A number of related issues remain open. Let us just mention a few of them. The unweighted norms of the HOPs are not yet determined in an explicit way for all $n$, nor in the extremal cases $n\rightarrow \infty$ and when the parameters of the weight function become large. The asymptotics ($q\rightarrow \infty$) of the unweighted norms for the Hermite and Laguerre polynomials is also unknown; indeed, a procedure not based on the Laplace formula is required as it was explained above. The explicit expression of the Shannon entropies of the HOPs in terms of the polynomial's degree and the parameters of the weight function has not yet been found, despite a recent effort \cite[Eq. (23)]{Toranzo2020} by means of some generalized hypergeometric functions evaluated at unity. Moreover, the asymptotics of the Shannon entropy of orthogonal polynomials in the whole Szeg\"{o} class is still unsolved; nevertheless, some remarkable results have been obtained \cite{Beckermann2004}. The calculation of the $\mathfrak{L}_{q}$-norms for the varying HOPs (i.e., polynomials whose weight-function's parameter does depend on the polynomial degree) and to discrete HOPs (Meixner, Hahn, Krawtchouk) is an open field to a great measure despite the publication of some interesting efforts (see e.g.\cite{Buyarov1999,Aptekarev2021bis,Levin2018,Sfetcu2016}). Finally, the extension of the discrete Shannon entropy of HOPs \cite{Aptekarev2009,Martinez2015} to the discrete $\mathfrak{L}_{q}$-norms has not yet been explored.

\section*{Acknowledgments}
This work has been partially supported by the Ministerio de Ciencia e Innovaci\'on (Spain) and the European Regional Development Fund (FEDER) under the grant PID2020-113390GB-I00.

\end{document}